\documentclass[dvips]{acta}
\usepackage{graphicx}
\usepackage{mathrsfs}
\usepackage{supertabular,lscape,epsfig}
\usepackage{subfigure}
\usepackage{amssymb}
\usepackage{amsmath}

\makeatletter
\newcommand{\cspace}{$\mathcal{C}$-space}
\newcommand{\chisq}{$\chi^2$}
\newcommand{\rchisq}{$\chi^2_{\rm red}$}
\makeatother
\begin{document}

\begin{Titlepage}
\Title{Principal Component Abundance Analysis of Microlensed Bulge Dwarf and
  Subgiant Stars}

\Author{B.~~H.~~A~n~d~r~e~w~s$^1$,~~~
D.~~H.~~W~e~i~n~b~e~r~g$^{1,2}$, ~~~
J.~~A.~~J~o~h~n~s~o~n$^{1,2}$,~~~
T.~~B~e~n~s~b~y$^3$,~~~and~~~
S.~~F~e~l~t~z~i~n~g$^3$}
{$^1$Department of Astronomy, The Ohio State University, 140 West 18th Avenue,
Columbus, OH 43210\\
email:(andrews, dhw, jaj)@astronomy.ohio-state.edu\\
$^2$Center for Cosmology and Astro-Particle Physics, The Ohio State University,
191 West Woodruff Avenue, Columbus, OH 43210\\
$^3$Lund Observatory, Department of Astronomy and Theoretical Physics, Box 43,
SE-22100 Lund, Sweden\\
e-mail:(tbensby, sofia)@astro.lu.se}

\Received{Month Day, Year}
\end{Titlepage}

\Abstract{Elemental abundance patterns can provide vital clues to the formation
and enrichment history of a stellar population.  Here we present an
investigation of the Galactic bulge, where we apply principal component
abundance analysis (PCAA)---a principal component decomposition of relative
abundances [$X$/Fe]---to a sample of 35 microlensed bulge dwarf and subgiant
stars, characterizing their distribution in the 12-dimensional space defined by
their measured elemental abundances.  The first principal component PC1, which
suffices to describe the abundance patterns of most stars in the sample, shows a
strong contribution from $\alpha$-elements, reflecting the relative
contributions of Type II and Type Ia supernovae.  The second principal component
PC2 is characterized by a Na--Ni correlation, the likely product of
metallicity-dependent Type II supernova yields.  The distribution in PC1 is
bimodal, showing that the bimodality previously found in the [Fe/H] values of
these stars is robustly and independently recovered by looking at only their
{\it relative} abundance patterns.  The two metal-rich stars that are
$\alpha$-enhanced have outlier values of PC2 and PC3, respectively, further
evidence that they have distinctive enrichment histories.  Applying PCAA to a
sample of local thin and thick disk dwarfs yields a nearly identical PC1; in
PC1, the metal-rich and metal-poor bulge dwarfs track kinematically selected
thin and thick disk dwarfs, respectively, suggesting broadly similar
$\alpha$-enrichment histories.  However, the disk PC2 is dominated by a Y--Ba
correlation, likely indicating a contribution of s-process enrichment from
long-lived asymptotic giant branch stars that is absent from the bulge PC2
because of its rapid formation.}{Galaxy: general --- Galaxy: bulge --- Galaxy:
evolution --- Galaxy: formation --- Galaxy: stellar content --- stars:
abundances}

\section{Introduction}
Elemental abundance trends of Galactic bulge stars are crucial for understanding
bulge formation.  Galactic chemical evolution is best traced by dwarf stars
because their spectra are straightforward to analyze (Edvardsson \etal 1993).
However, observations of bulge dwarfs are challenging due to their faintness
($V$=19--20; Feltzing and Gilmore 2000), impeding spectroscopic observation
under normal circumstances.  Consequently, many studies have focused on giant
stars, despite the difficulty in analyzing their spectra.  This difficulty has
led to shifts in the mean [Fe/H] and the [Fe/H] distribution function of the
bulge as the analysis techniques are refined (\eg Rich 1988, McWilliam and Rich
1994, Fulbright \etal 2007, Zoccali \etal 2008, Hill \etal 2011).  Fortunately,
gravitational microlensing offers a unique opportunity to observe bulge dwarfs.
When a bulge dwarf is lensed by a foreground object, its brightness can increase
by $>$5 magnitudes, enabling spectroscopic observations of sufficiently high
resolution and signal-to-noise (S/N) for an abundance analysis (\eg Minniti
\etal 1998, Johnson \etal 2007, Bensby \etal 2011 and references therein).

The abundances of different elements are correlated, reflecting their origin in
a common nucleosynthetic process, such as Type II or Type Ia supernovae (SNe).
However, these correlations are not perfect because the same element can be made
in multiple nucleosynthetic processes, whose relative contributions to a star
are best distinguished if abundances are measured for many elements.  Here we
analyze a sample of 35 bulge dwarfs\footnote{Although our sample includes some
subgiant stars, we will describe it as ``bulge dwarfs'' for brevity.}, all of
which have at least seven measured elemental abundances and the majority (24/35)
of which have 11 or 12 measured elemental abundances, with median errors $\sigma
< 0.25$ dex.  Bensby \etal (2011) find that the bulge dwarf [Fe/H] distribution
function is bimodal, peaked at [Fe/H] $\approx -0.6 \; {\rm and} \; +0.3$.  They
further show that the $\alpha$-element abundances in the bulge dwarfs vary
systematically with [Fe/H], following the trends found for thin and thick disk
dwarfs (Bensby \etal 2003, 2005, Reddy \etal 2003, 2006).  Here we revisit this
data set with a different analysis technique based on principal component
decomposition of the elemental abundance patterns, showing that the bimodality
seen in [Fe/H] also appears in the {\it relative} elemental abundance patterns.

Principal component analysis (PCA) is a natural tool for characterizing
correlations in a high-dimensional space, reducing the overall dimensionality of
the data set while allowing the data themselves to reveal the strongest patterns
of correlations.  While PCA has a long history in astronomy, its application to
elemental abundance analysis is relatively new.  The only such application we
know of is the study of Ting \etal (2012), who used this technique to
investigate the distributions in elemental abundance space (hereafter, \cspace;
Freeman and Bland-Hawthorn 2002) defined by various samples of stars from the
disk, halo, clusters, and satellite galaxies.  They found that disk stars
occupied about 6 dimensions within the 17-dimensional \cspace\ of the data, but
the nucleosynthetic processes likely responsible for the lowest order dimensions
changed as a function of [Fe/H].  Their work demonstrated the potential
usefulness of principal component abundance analysis (PCAA) as a way to identify
groups of stars with distinct enrichment histories.  Here we apply this approach
to bulge dwarfs to shed further light on the formation of the Galactic bulge.


\section{Method}
PCAA defines a new set of orthogonal basis vectors in \cspace\ whose components
are chosen to align with the maximum variation within the data not already
attributed to lower order components.  We use standard PCA (see, \eg Jolliffe
1986) with the data matrix \{$d_{i,j}$\} representing the logarithmic
abundance\footnote{Elemental abundances are defined as [$X/Y$] $\equiv$ ${\rm
log}(N_X / N_Y) - {\rm log}(N_X / N_Y)_\odot$, with missing data replaced by the
average value of that elemental abundance for the other stars.}  relative to
iron of element $j$ for star $i$: $d_{i,j}$ = [$X_j$/Fe] with $X_j$ = O, Na, Mg,
Al, Si, Ca, Ti, Cr, Ni, Zn, Y, and Ba for $j$ = 1, 2, ..., 12.  PCA identifies
orthogonal eigenvectors ${\bf e}_{k} = \{e_{k,j}\}$ such that the abundance of a
given element in a given star can be represented as a sum
\begin{equation}
\label{eqn:pcbasis}
d_{i,j} = \bar{d_j} + \displaystyle\sum\limits_{k=1}^{N_{\rm PC}} c_{i,k}
e_{k,j},
\end{equation}
where $\bar{d_{j}} = \frac{1}{N_{\rm star}}
\displaystyle\sum\limits_{i=1}^{N_{\rm star}} d_{i,j}$ is the mean value of
[$X_j$/Fe] in the full data set and $c_{i,k}$ is the coefficient for the $k^{\rm
th}$ PC of star $i$.  The first PC describes the direction in elemental
abundance space along which the sample stars exhibit the greatest variation, the
second PC describes the direction of the second largest variation, etc. If the
number of principal components in the sum is equal to the number of elements
measured, then the data can be represented exactly.  However, if elemental
abundances are correlated so that stars are restricted to a lower dimensional
subspace, then the elemental abundances can be represented to good accuracy by a
smaller number of PCs.

Our data set comes from the homogeneous elemental abundance and error reanalysis
of microlensed bulge dwarfs and subgiants by Bensby et al.~(in prep.);
abundances for 26 of our 35 sample stars have been previously published (Epstein
\etal 2010, Bensby \etal 2010, 2011).


\section{Results}
The dimensionality of the subspace occupied by stars within the full \cspace\
can be expressed as the number of PCs required to explain the intrinsic
variation in the data (\ie the variation not attributable to observational
errors).  Ting \etal (2012) used Monte Carlo simulations spanning a range of
intrinsic dimensionality and variance to show that the true dimensionality was
recovered when the cumulative variation of the first $k$ PCs was about 85\%.
For our sample of microlensed bulge dwarfs, the first 1, 2, and 3 PCs describe
64\%, 77\%, and 84\% of the cumulative variation within the data.  Although the
threshold for completely describing the data likely depends on many factors
including sample size, number of observed abundances, and abundance
uncertainties, it is fair to say that the bulge dwarfs occupy approximately
three dimensions of the 12-dimensional \cspace\ investigated here.



\begin{figure}[htb]
\mbox{\subfigure{\includegraphics[width=2.5in]{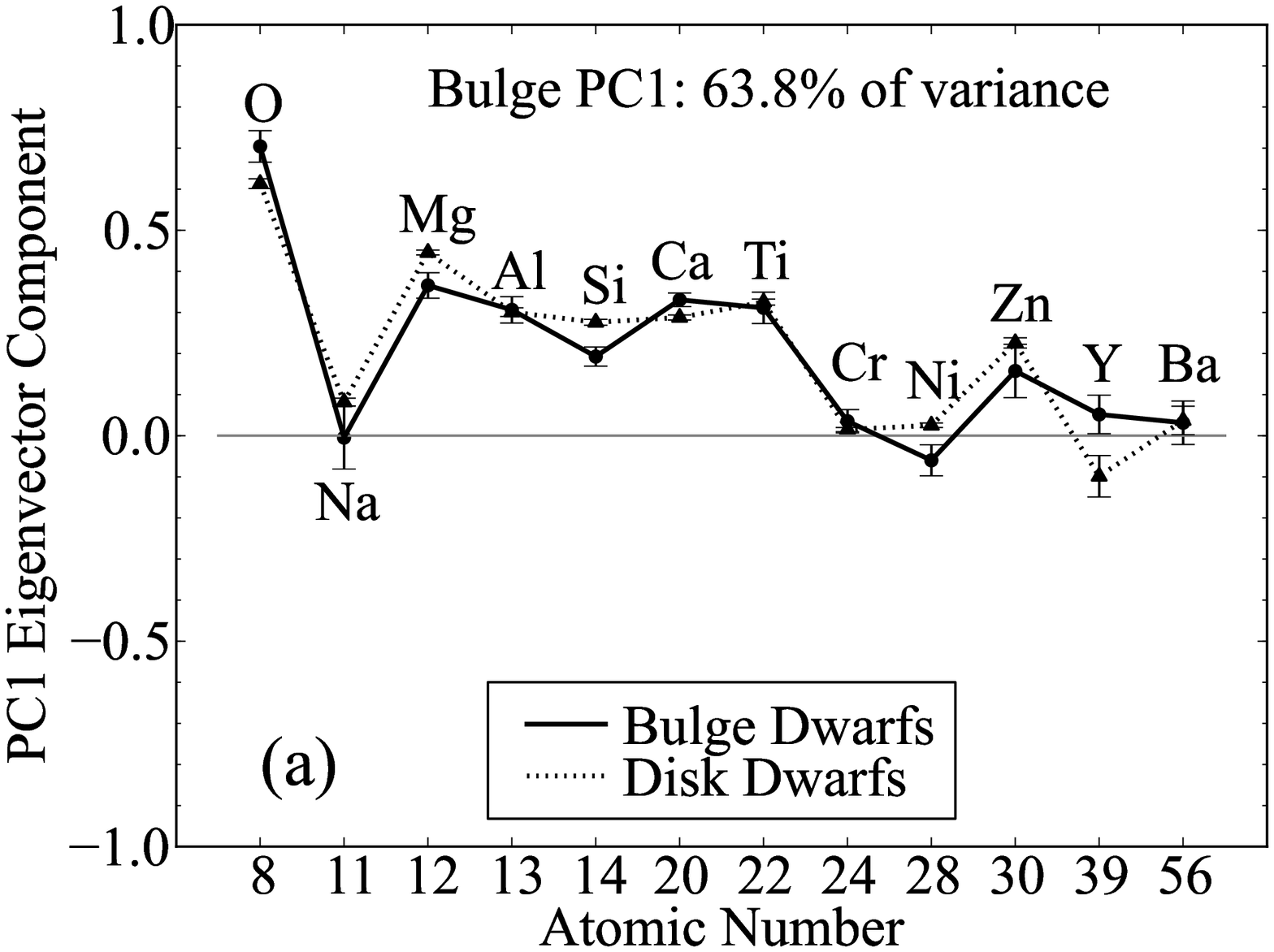}}\;
\subfigure{\includegraphics[width=2.5in]{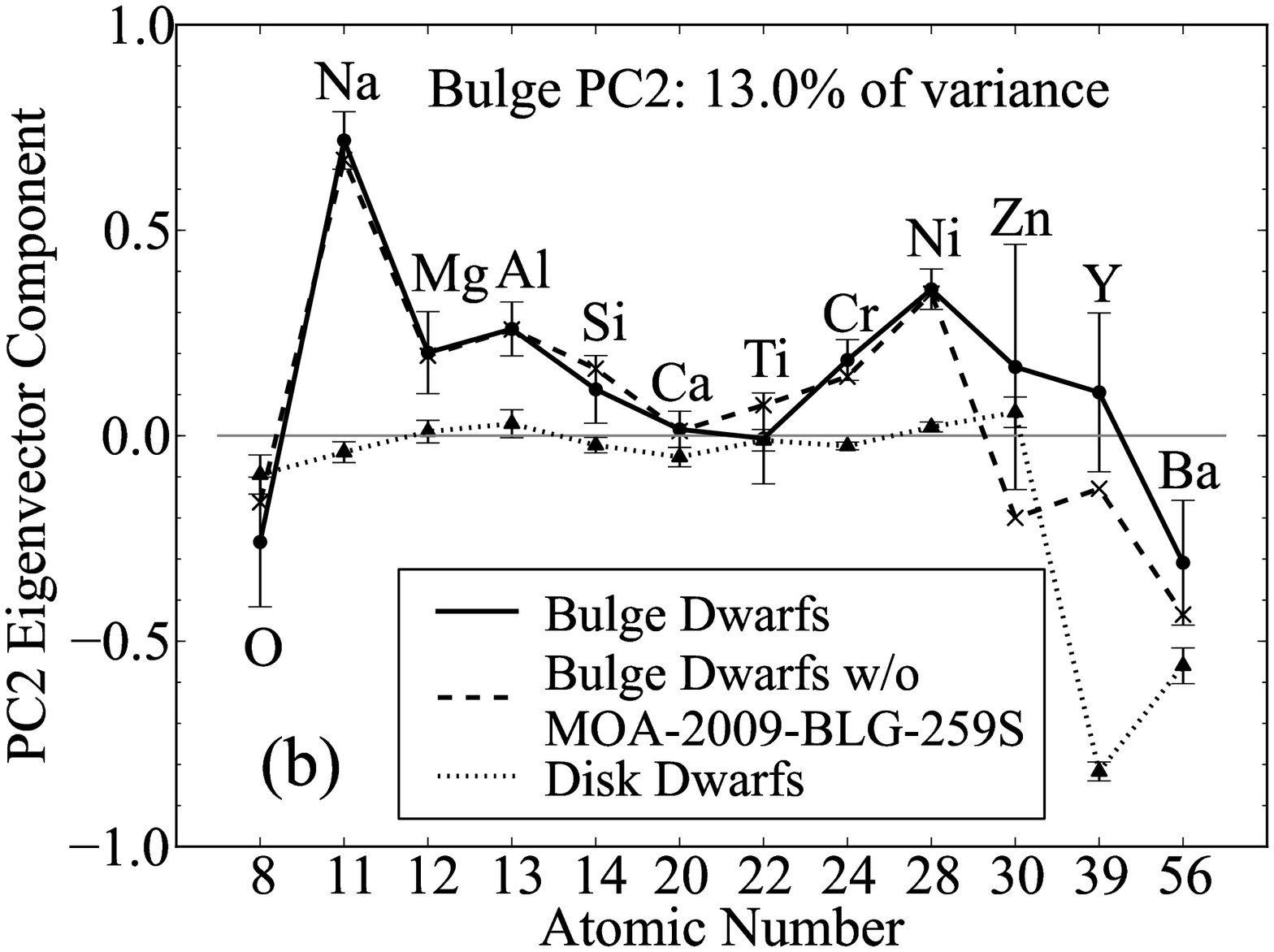}}}
\FigCap{A graphical representation of the PC1 (panel a) and PC2 (panel b)
vectors of the microlensed bulge dwarfs (circles/solid lines) and the disk dwarf
sample of Bensby \etal (2003, 2005, and in prep.) (triangles/dotted lines).  The
PC1 and PC2 eigenvector components refer to $e_{1,j}$ and $e_{2,j}$,
respectively, from Eq.~(1), with uncertainties from bootstrap resampling.  We
also show PC2 for the bulge dwarfs if MOA-2009-BLG-259S is omitted from the
sample (crosses/dashed line in panel b); PC1 remains nearly unchanged, so it is
not plotted for clarity.  Each symbol represents an abundance ([$X$/Fe]) and
thus a dimension in the original \cspace.}
\end{figure}

Fig.~1 shows the first two PCs derived from the observed bulge dwarf elemental
abundances, with the uncertainties determined from bootstrap resampling (see
below).  PC1 is dominated by the abundances of oxygen, other $\alpha$-elements
(Mg, Si, Ca, and Ti), and Al, with small uncertainties on the relative
contributions of each abundance.  SNe II are the primary source of
$\alpha$-elements and Al, but they are also a significant producer of Fe,
especially at early times.  By contrast, SNe Ia create large amounts of Fe and
other Fe-peak elements, leading to sub-solar [$\alpha$/Fe] yields; once enough
time has elapsed for a substantial number of SNe Ia to occur, they become the
dominant source of Fe.  The interplay between these two nucleosynthetic sources
is thought to underpin the dichotomy in [$\alpha$/Fe] observed in Galactic disk
stars, with high [$\alpha$/Fe] for ``thick disk'' stars reflecting rapid
formation (\eg Fuhrmann 1998) and roughly solar [$\alpha$/Fe] for ``thin disk''
stars, which have predominantly higher [Fe/H] (Gilmore and Wyse 1985).  Although
PCA is a ``blind'' statistical technique with no {\it a priori} theoretical
input, in this data set (and others we have explored) the first principal
component picks up this expected distinction between the two dominant supernova
enrichment mechanisms.

PC2 is primarily governed by Na with secondary contributions from Ni (correlated
with Na) and Ba (anticorrelated with Na).  A similar Na--Ni correlation,
attributed to metallicity-dependent SN II yields, has been found previously in
halo stars (Nissen and Schuster 1997, 2010); however, this study and the
companion paper of Bensby et al.~(in prep.) are the first to identify a Na--Ni
correlation amongst bulge stars (see Bensby et al.~in prep.~for more details).
Na is primarily produced by hydrostatic carbon burning in the massive stars that
explode as SNe II, but proton capture at the same temperatures depletes the
pre-explosion Na abundance.  As metallicity increases, the neutron excess
increases, making Na less susceptible to proton capture and consequently
increasing the overall Na yield (Clayton 2003).  Similarly, the yield of
$^{58}$Ni (the most common Ni isotope) from SNe II is sensitive to the neutron
excess and the abundance of neutron-rich nuclei, like $^{23}$Na, in the
progenitor star (Woosley \etal 1973).  Additional significant $^{58}$Ni
production occurs in SNe Ia, whose $^{58}$Ni yield increases with metallicity
(Timmes \etal 2003).  However, the Na--Ba anticorrelation is not readily
explained by a single nucleosynthetic process.  One star with distinctive
abundances, MOA-2009-BLG-259S (see Fig.~3c), has a large impact on the
contributions of Zn, Y, and Ba to PC2 and drives up the uncertainties for these
abundances.  The crosses/dashed line in Fig.~1b show the effect of omitting this
star when defining principal components; PC1 hardly changes, but the
contributions of Zn and Y to PC2 switch from being correlated with Na to
anticorrelated.  The [Y/Fe] for MOA-2009-BLG-259S has a large uncertainty,
though the elevated ${\rm [Zn/Fe]} = 0.44 \pm 0.17$ appears to be
well-established (Fig.~3c).  Regardless of MOA-2009-BLG-259S, PC2 is dominated
by Na and shows a Na--Ni correlation and a Na--Ba anticorrelation.

For comparison, we have found the principal components of a sample of 702 solar
neighborhood thin and thick disk dwarfs from Bensby \etal (2003, 2005, and in
prep.)  with the same elements measured.  The disk PC1 and PC2 are shown as
triangles/dotted lines in Fig.~1.  The clear similarity between the bulge and
disk PC1s suggests that the relative enrichment from SNe II vs.~SNe Ia is the
main driver of diversity among stars in both samples.  On the other hand, the
disk and bulge PC2s do not resemble each other: in contrast to the bulge PC2
discussed above, the disk PC2 is dominated entirely by correlated Y and Ba, both
neutron-capture elements produced mainly by the $s$-process in disk stars
(Sneden \etal 2008).  The short vs.~long lifetimes of the nucleosynthetic
sources driving the bulge and disk PC2s, respectively, implies that the bulge
formed too rapidly ($\lesssim$Gyr) for asymptotic giant branch stars to dominate
PC2.  Johnson \etal (2012) independently reached a similar conclusion based on
the relative abundances of $r$- and $s$-process elements.



\begin{figure}[htb]
\includegraphics[width=12cm]{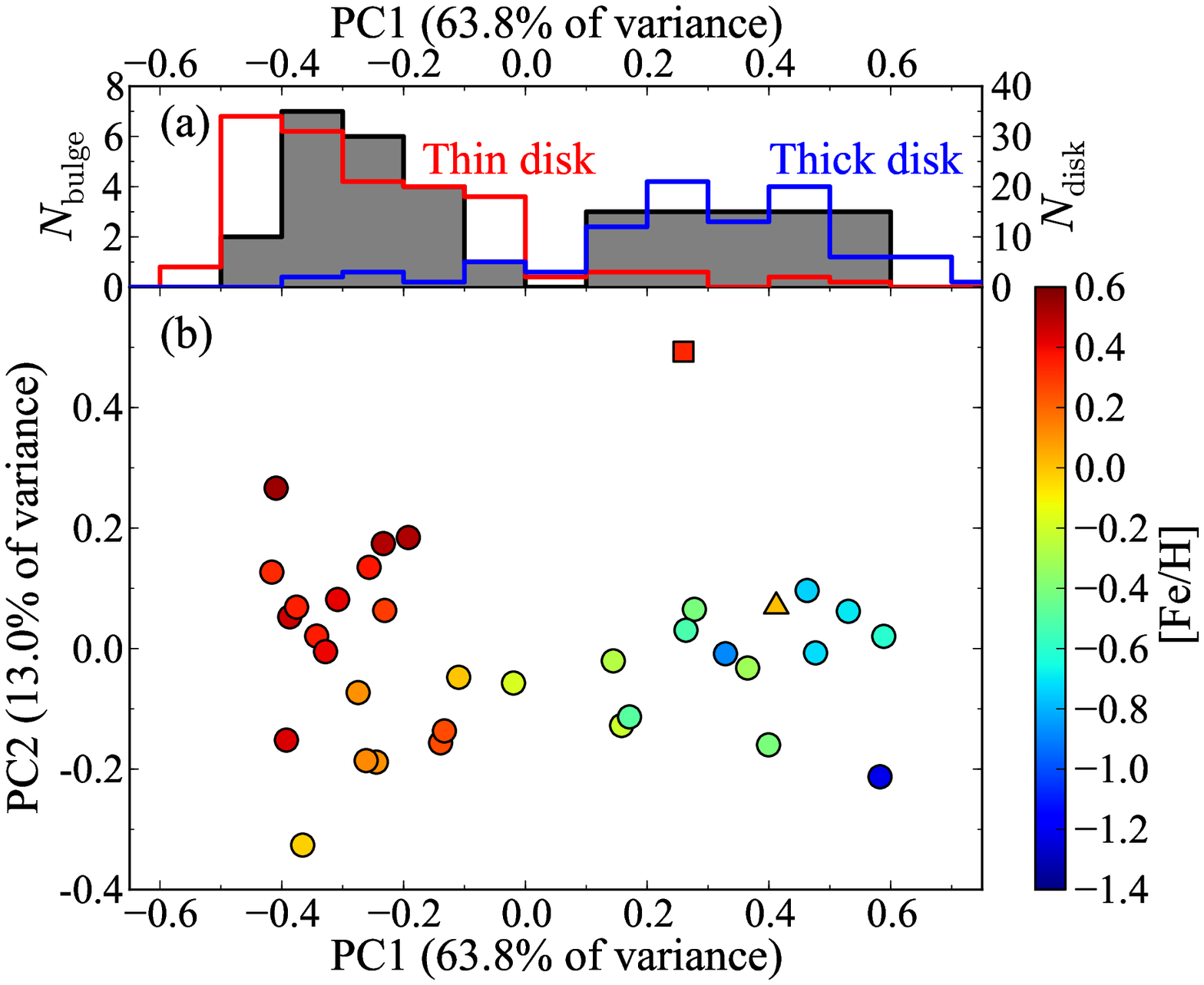}
\FigCap{Panel (a): the histograms of bulge
dwarfs (gray) and of kinematically selected thin (red) and thick (blue) disk
dwarfs projected onto the bulge PC1 axis defined above.  Panel (b): the
distribution of bulge dwarfs in (PC1, PC2)-space, color-coded by [Fe/H].  The
square and triangle represent the PC2 outlier (MOA-2009-BLG-259S) and the PC3
outlier (MOA-2010-BLG-523S), respectively.  The positions along PC1 and PC2
correspond to $c_{i,1}$ and $c_{i,2}$, respectively, in Eq.~(1).}
\end{figure}

Fig.~2 shows the distribution of the 35 bulge dwarfs in (PC1, PC2)-space, with
the upper panel showing the histograms of the bulge dwarfs in PC1 and of
kinematically selected subsets of thin and thick disk dwarfs\footnote{The
kinematic designation is based on the Bensby \etal (2003) selection criteria;
however, we adopt a more stringent cut on the relative thick/thin disk
membership probability: stars with $P$(thick disk)/$P$(thin disk) $>$ 100 and
$P$(thick disk)/$P$(thin disk) $<$ 0.01 are designated thick and thin disk
stars, respectively.} projected onto the bulge PC1.  It is evident from visual
inspection that the bulge dwarfs divide into two distinct groups along the PC1
axis, centered at PC1 values of $\pm0.3$.  (Because the sum in Eq.~1 includes
the mean elemental abundances of the sample, a value of ${\rm PC1} = -0.3$
corresponds to $\sim$[$\alpha$/Fe]$_\odot$.)  Thus, this analysis of {\it
relative} elemental abundances, with no direct input from [Fe/H], recovers the
bimodality that Bensby \etal (2011) found in the [Fe/H] distribution without
reference to relative abundances.  Bensby \etal (2011) did find elevated
[$\alpha$/Fe] ratios for the metal-poor bulge dwarfs, and we recover the same
correlation in this ``reverse'' analysis: every star with ${\rm PC1} < -0.1$ has
${\rm [Fe/H]} > -0.02$, and all but two of the stars with ${\rm PC1} > -0.1$
have ${\rm [Fe/H]} < -0.18$.  Of these two stars, one (MOA-2009-BLG-259S;
square) is a clear PC2 outlier, while the other (MOA-2010-BLG-523S; triangle) is
a moderate PC3 outlier that is undistinguished in (PC1, PC2)-space.

Fig.~2a shows that the metal-rich and metal-poor bulge dwarfs track the thin and
thick disk dwarfs, respectively, in PC1.  PCAA highlights the scarcity of stars
with intermediate [$\alpha$/Fe] (PC1 $\sim$ 0) in the bulge dwarfs and in the
thin and thick disk dwarfs.  The fact that the two bulge populations track the
two disk populations in [Fe/H] and PC1 suggests that the stars had
$\alpha$-enrichment histories that produced the same abundance patterns.  This
similarity provides tentative evidence that the bulge formed through secular
processes, such as disk instabilities, that heated inner thin and thick disk
stars to form the bulge (Kormendy and Kennicutt 2004).  Alternatively, the bulge
and disk could have formed by distinct mechanisms but experienced ``convergent''
enrichment histories.  For example, Bournaud \etal (2007) propose that the large
cosmological accretion rates of high redshift galaxies enable the rapid
($\sim$0.5--1 Gyr), simultaneous formation of the bulge and inner disk, which
could account for the $\alpha$-enhanced subpopulation of each component.

To test the statistical significance of PC1 and PC2 and the robustness of the
bimodality in Fig.~2, we created 100 bootstrap resamplings of the data set,
redefining principal components each time.  PC1 is always recovered, and the
general form of PC2 (high Na value, Na--Ni correlation, and Na--Ba
anticorrelation) is recovered in 99/100 resamplings.  Thus, PC2 is statistically
significant even though the contributions of Zn, Y, and Ba to PC2 fluctuate
because of its sensitivity to MOA-2009-BLG-259S.  The histogram of PC1 values is
always multimodal, showing two distinct groupings (as in Fig.~2a) about 90\% of
the time; the remaining resamplings show three apparent groupings, but the small
sample size makes the distribution of values {\it within} the ${\rm PC1} > 0$
group difficult to characterize reliably.  The significance of a formal test for
bimodality will depend on the form of the adopted null hypothesis, but the
likelihood ratio of a 2-Gaussian fit to the PC1 distribution to a 1-Gaussian fit
is $\mathscr{L}(2{\rm G})/\mathscr{L}(1{\rm G}) = 2 \times 10^5$, a large
improvement for the addition of two free parameters.



\begin{figure}[htb]
\includegraphics[width=12cm]{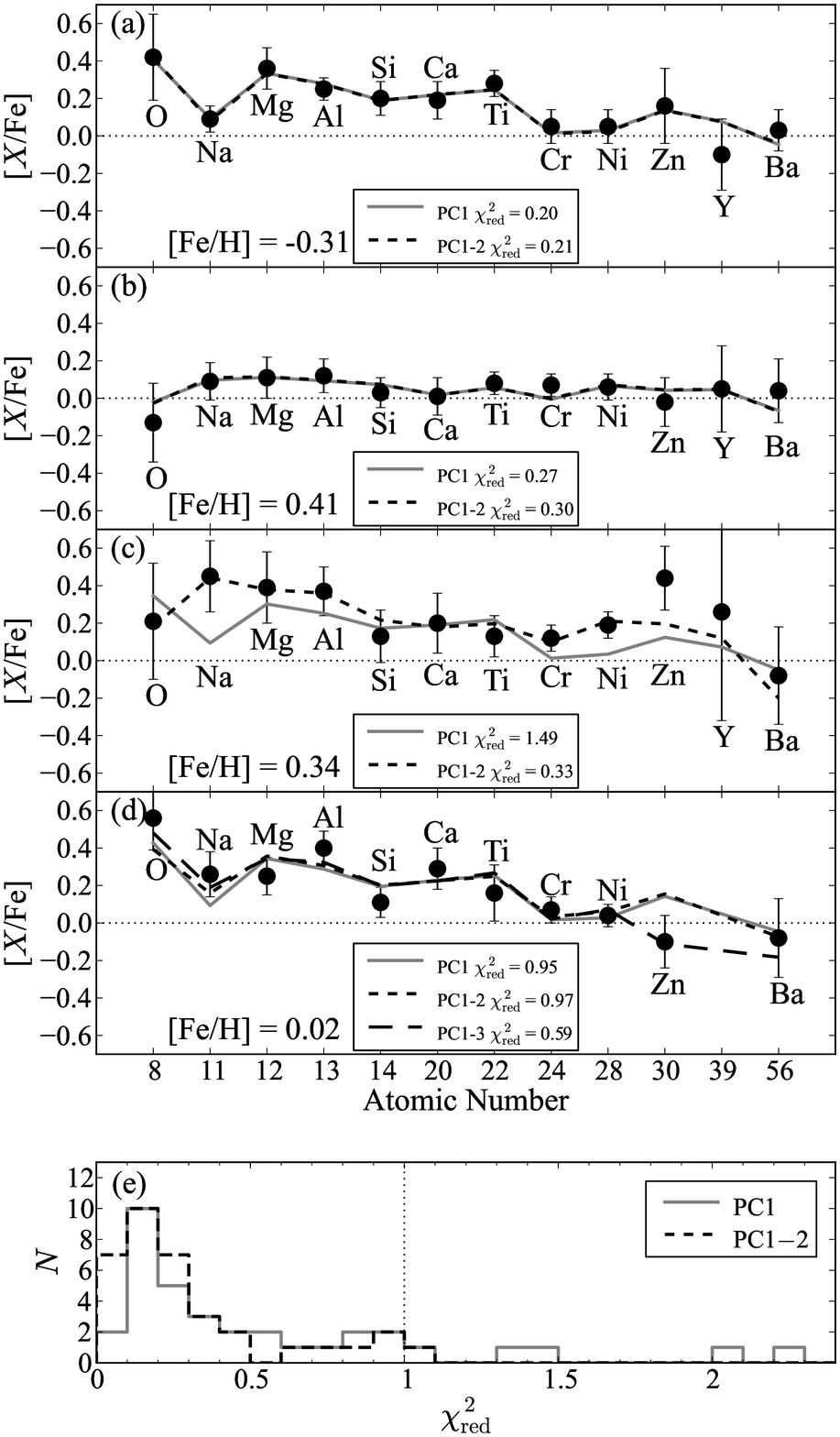}
\FigCap{Panels (a)--(d) show elemental abundances for, respectively, a
  metal-poor star, a metal-rich star, the PC2 outlier star (MOA-2009-BLG-259S),
  and the PC3 outlier star (MOA-2010-BLG-523S).  Lines indicate the best fit
  combined abundance pattern using the mean abundance and the first PC (solid
  gray line), the first two PCs (short dashed line), and the first three PCs
  (long dashed line in panel d only).  Reduced-\chisq\ (\rchisq) values for
  these fits are listed in the legends.  Panel (e) shows the distribution of
  \rchisq\ values for the bulge dwarf sample, with the same line types as panels
  (a)--(c).}
\end{figure}

Fig.~3 shows the decompositions of the elemental abundance patterns of four
sample stars into the sum of the sample mean and the first one, two, or three
PCs.  The best fit coefficients $c_{i, k}$ of Eq.~(1) are found for each star by
\chisq-minimization, treating all errors as independent.  Panels (a) and (b)
show typical examples of metal-poor and metal-rich stars, respectively, each of
them fit almost perfectly with a single PC.  Panel (c) shows MOA-2009-BLG-259S,
which is poorly fit by PC1 alone but well fit when PC2 is also included.  Panel
(d) shows MOA-2010-BLG-523S, which has one of the largest PC3 coefficients in
our sample.

Clearly these PCA fits have low values of \rchisq\ = \chisq/(d.o.f.), once a
sufficient number of PCs (one, two, or three) is included.  The number of
degrees of freedom is the number of elemental abundances measured for the star
minus the number of PCs in the fit.  Panel (e) shows the histogram of \rchisq\
values for single-PC and 2-PC decompositions; median values are \rchisq\ = 0.31
and 0.21, respectively.  The low \rchisq\ values indicate that the observational
errors on the elemental abundances are typically overestimated, at least in the
sense that they do not represent the variance in estimated elemental abundance
that would arise from observing the same star many times.  We believe that these
low \rchisq\ values arise because the quoted errors effectively incorporate a
component of systematic calibration error in addition to random error.  For
example, multiple lines of the same element may give discrepant abundance
estimates because of uncertainties in the assumed oscillator strengths; the
internal dispersion of these estimates is one useful indication of the absolute
elemental abundance uncertainty, but the range of elemental abundance
measurements from multiple high-S/N spectra will be smaller than this dispersion
because the same oscillator strengths are assumed each time.  The treatment of
random and systematic errors and error correlations is a significant issue for
PCAA and other model-fitting approaches, but we defer it to future
investigations.


\section{Conclusions}
Our results confirm and extend the findings of Bensby \etal (2011), who
identified bimodality in the [Fe/H] distribution of microlensed bulge dwarfs,
with metal-poor stars showing enhanced $\alpha$-element abundances like those of
solar neighborhood thick disk stars.  Our principal component analysis
demonstrates that the bimodality found by Bensby \etal (2011) is present even in
the {\it relative} elemental abundances ([$X$/Fe]) of the bulge dwarfs alone.
The first PC is dominated by $\alpha$-elements, reflecting the dichotomy between
SN II and SN Ia enrichment.  The second PC captures the Na--Ni correlation
caused by the metallicity dependence of SN II yields.  Intriguingly, the two
metal-rich stars that exhibit $\alpha$-enhancement (MOA-2009-BLG-259S and
MOA-2010-BLG-523S) are also outliers from the main locus of stars in (PC1, PC2,
PC3)-space, suggesting that they do indeed have unusual enrichment histories.
If we project the thin and thick disk dwarfs onto the bulge dwarf PC1, they
occupy the locations of the metal-rich and metal-poor bulge dwarfs,
respectively.  Analyzing local disk dwarfs, we find that the first principal
component is nearly identical to the bulge PC1.  However, the disk PC2 is
governed by Y and Ba, products of long-lived asymptotic giant branch stars,
whereas enrichment from short-lived SNe II drives the bulge PC2.  Qualitatively,
these results support a scenario in which the bulge grows by secular evolution
from the inner disk, which itself has the elemental abundance dichotomy seen in
local populations, but there may be other bulge formation models that can
produce similar results.

Our results, and those of Ting \etal (2012), illustrate the potential of PCAA as
a tool for characterizing the distribution of stars in high-dimensional
\cspace.  One application, highlighted here, is to identify subpopulations in a
sample, drawing on the information present in all measured elemental abundances
simultaneously.  With large multi-element samples, this approach could be used
to isolate ``interloper'' stars accreted from a dissolved satellite, and perhaps
to identify cohorts of stars associated with common birth clusters (Freeman and
Bland-Hawthorn 2002).  A second application, illustrated by the examples of
MOA-2009-BLG-259S and MOA-2010-BLG-523S, is to identify outlier stars, either
through their unusual locations in PC-space or because they are poorly fit by
combinations of PCs that fit most stars well.  Such outliers may reveal rare but
physically informative enrichment pathways.  A third application, highlighted by
Ting \etal (2012), is to characterize the dimensionality of the stellar
distribution in \cspace, which is a basic test for models of Galactic assembly
and enrichment.  We will explore this technique in future work using theoretical
models.

By allowing the data themselves to define the directions of strongest variation,
PCAA complements the usual approach of testing predictive models that adopt
nucleosynthetic yields from theoretical calculations.  The PCs clearly do have
physical content, but the connection of PCs to enrichment mechanisms is not
one-to-one (see Ting \etal 2012), and interpreting them will require comparisons
to models that vary both enrichment and mixing histories and the nucleosynthetic
yields themselves.  Our analysis identifies several practical complications of
PCAA, including the potential sensitivity to outliers in small samples, the
treatment of missing elemental abundance measurements, the impact of
heteroscedastic errors and correlated errors, and the mix of random and
systematic contributions to the errors quoted in observational analyses.  We
will investigate these issues in future work.  The enormous samples of
high-resolution spectra anticipated from the SDSS-III APOGEE survey (Majewski
\etal in prep., Eisenstein \etal 2011), the Gaia-ESO survey (Gilmore \etal
2012), and the HERMES survey (Barden \etal 2010) will map the elemental
abundance distribution over wide swaths of the Galaxy, and PCAA will be a
valuable tool for connecting these measurements to a comprehensive theory of the
formation of the Milky Way.


\Acknow{We acknowledge support of NSF grant AST1009505.  T.B. was funded by
grant No.~621-2009-3911 from The Swedish Research Council.}


\begin {references} 

\refitem{Barden, S.~C., Jones, D.~J., Barnes, S.~I., \etal}{2010}{Proc.~SPIE}{7735}{}

\refitem{Bensby, T., Feltzing, S., and Lundstr{\"o}m, I.}{2003}{\AA}{410}{527}

\refitem{Bensby, T., Feltzing, S., Lundstr{\"o}m, I., and Ilyin, I.}{2005}{\AA}{433}{185}

\refitem{Bensby, T., \etal}{2010}{\AA}{512}{A41+}

\refitem{Bensby, T., \etal}{2011}{\AA}{533}{A134}

\refitem{Bournaud, F., Elmegreen, B.~G., and Elmegreen, D.~M.}{2007}{\ApJ}{670}{237}

\refitem{Clayton, D.}{2003}{Handbook of Isotopes in the Cosmos, by Donald
  Clayton, pp.~326.~ISBN 0521823811.~Cambridge, UK: Cambridge University Press,
  October 2003.}{}{}

\refitem{Edvardsson, B., Andersen, J., Gustafsson, B., Lambert, D.~L., Nissen,
  P.~E., and Tomkin, J.}{1993}{\AA}{275}{101}

\refitem{Eisenstein, D.~J., \etal}{2011}{\AJ}{142}{72}

\refitem{Epstein, C.~R., Johnson, J.~A., Dong, S., Udalski, A., Gould, A., and
  Becker, G.}{2010}{\ApJ}{709}{447}

\refitem{Feltzing, S., and Gilmore, G.}{2000}{\AA}{355}{949}

\refitem{Freeman, K., and Bland-Hawthorn, J.}{2002}{ARA\&A}{40}{487}

\refitem{Fuhrmann, K.}{1998}{\AA}{338}{161}

\refitem{Fulbright, J.~P., McWilliam, A., and Rich, R.~M.}{2007}{\ApJ}{661}{1152}

\refitem{Gilmore, G., and Wyse, R.~F.~G.}{1985}{\AJ}{90}{2015}

\refitem{Gilmore, G., \etal}{2012}{The Messenger}{147}{25}

\refitem{Hill, V., \etal}{2011}{\AA}{534}{A80}

\refitem{Johnson, C.~I., Rich, R.~M., Kobayashi, C., and Fulbright, J.~P.}{2012}{\ApJ}{749}{175}

\refitem{Johnson, J.~A., Gal-Yam, A., Leonard, D.~C., Simon, J.~D., Udalski,
  A., and Gould, A.}{2007}{\ApJ}{655}{L33}

\refitem{Jolliffe, I.~T.}{1986}{Springer Series in Statistics, Berlin: Springer,
1986}{}{}

\refitem{Kormendy, J., and Kennicutt, Jr., R.~C.}{2004}{ARA\&A}{42}{603}

\refitem{McWilliam, A., and Rich, R.~M.}{1994}{\ApJS}{91}{749}

\refitem{Minniti, D., Vandehei, T., Cook, K.~H., Griest, K., and Alcock, C.}{1998}{\ApJ}{499}{L175}

\refitem{Nissen, P.~E., and Schuster, W.~J.}{1997}{\AA}{326}{751}

\refitem{Nissen, P.~E., and Schuster, W.~J.}{2010}{\AA}{511}{L10}

\refitem{Reddy, B.~E., Lambert, D.~L., and Allende Prieto, C.}{2006}{\MNRAS}{367}{1329}

\refitem{Reddy, B.~E., Tomkin, J., Lambert, D.~L., and Allende Prieto, C.}{2003}{\MNRAS}{340}{304}

\refitem{Rich, R.~M.}{1988}{\AJ}{95}{828}

\refitem{Sneden, C., Cowan, J.~J., and Gallino, R.}{2008}{ARA\&A}{46}{241}

\refitem{Timmes, F.~X., Brown, E.~F., and Truran, J.~W.}{2003}{\ApJ}{590}{L83}

\refitem{Ting, Y.~S., Freeman, K.~C., Kobayashi, C., de Silva, G.~M., and Bland-Hawthorn, J.}{2012}{\MNRAS}{421}{1231}

\refitem{Woosley, S.~E., Arnett, W.~D., and Clayton, D.~D.}{1973}{\ApJS}{26}{231}

\refitem{Zoccali, M., Hill, V., Lecureur, A., Barbuy, B., Renzini, A., Minniti, D., G{\'o}mez, A., and Ortolani, S.}{2008}{\AA}{486}{177}

\end{references}
\end{document}